\documentclass[aps,prl,reprint,superscriptaddress,amsmath,amssymb,longbibliography]{revtex4-2}

\usepackage{graphicx}
\usepackage{longtable}
\usepackage{xcolor}
\usepackage[bookmarks,bookmarksnumbered,colorlinks,allcolors=blue]{hyperref}
\usepackage{bookmark}
\usepackage{ifthen}
\usepackage{ulem}
\usepackage{physics}

\definecolor{dkgreen}{rgb}{0,0.6,0}
\definecolor{gray}{rgb}{0.5,0.5,0.5}
\definecolor{mauve}{rgb}{0.58,0,0.82}

\newcommand{\MHz}{\ensuremath{\mathrm{MHz}}}
\newcommand{\GHz}{\ensuremath{\mathrm{GHz}}}

\newcommand{\insitu}{\textit{in situ}\ }
\def\be#1\ee{\begin{equation}#1\end{equation}}
\def\ba#1\ea{\begin{align}#1\end{align}}
\def\bg#1\eg{\begin{gather}#1\end{gather}}

\newcommand{\Teff}{T_{\text{eff}}}
\newcommand{\phiext}{\Phi_{\text{ext}}}

\begin{document}

\title{Characterization of loss mechanisms in a fluxonium qubit}

\author{Hantao Sun}
\thanks{These two authors contributed equally}
\affiliation{Alibaba Quantum Laboratory, Alibaba Group, Hangzhou, Zhejiang 311121, China}
\author{Feng Wu}
\thanks{These two authors contributed equally}
\affiliation{Alibaba Quantum Laboratory, Alibaba Group, Hangzhou, Zhejiang 311121, China}
\author{Hsiang-Sheng Ku}
\affiliation{Alibaba Quantum Laboratory, Alibaba Group, Hangzhou, Zhejiang 311121, China}
\author{Xizheng Ma}
\affiliation{Alibaba Quantum Laboratory, Alibaba Group, Hangzhou, Zhejiang 311121, China}
\author{Jin Qin}
\affiliation{Alibaba Quantum Laboratory, Alibaba Group, Hangzhou, Zhejiang 311121, China}
\author{Zhijun Song}
\affiliation{Alibaba Quantum Laboratory, Alibaba Group, Hangzhou, Zhejiang 311121, China}
\author{Tenghui Wang}
\affiliation{Alibaba Quantum Laboratory, Alibaba Group, Hangzhou, Zhejiang 311121, China}
\author{Gengyan Zhang}
\affiliation{Alibaba Quantum Laboratory, Alibaba Group, Hangzhou, Zhejiang 311121, China}
\author{Jingwei Zhou}
\affiliation{Alibaba Quantum Laboratory, Alibaba Group, Hangzhou, Zhejiang 311121, China}
\author{Yaoyun Shi}
\affiliation{Alibaba Quantum Laboratory, Alibaba Group USA, Bellevue, WA 98004, USA}
\author{Hui-Hai Zhao}
\email{huihai.zhh@alibaba-inc.com}
\affiliation{Alibaba Quantum Laboratory, Alibaba Group, Beijing 100102, China}
\author{Chunqing Deng}
\email{chunqing.cd@alibaba-inc.com}
\affiliation{Alibaba Quantum Laboratory, Alibaba Group, Hangzhou, Zhejiang 311121, China}

\begin{abstract}
Using a fluxonium qubit with \insitu tunability of its Josephson energy, we characterize its energy relaxation at different flux biases as well as different Josephson energy values. The relaxation rate at qubit energy values, ranging more than one order of magnitude around the thermal energy $k_B T$, can be quantitatively explained by a combination of dielectric loss and $1/f$ flux noise with a crossover point. The amplitude of the $1/f$ flux noise is consistent with that extracted from the qubit dephasing measurements at the flux sensitive points. In the dielectric loss dominant regime, the loss is consistent with that arises from the electric dipole interaction with two-level-system (TLS) defects. In particular, as increasing Josephson energy thus decreasing qubit frequency at the flux insensitive spot, we find that the qubit exhibits increasingly weaker coupling to TLS defects thus desirable for high-fidelity quantum operations. 
\end{abstract}
\maketitle

\bookmarksetup{startatroot}

Fluxonium is a type of superconducting qubits that feature large anharmonicity~\cite{manucharyan2009fluxonium} and high coherence~\cite{nguyen2019high}, therefore attractive for high-fidelity quantum computing systems~\cite{ficheux2021fast, bao2022fluxonium, moskalenko2022high, dogan2022demonstration, nguyen2022blueprint}. Due to the rich structure of its energy levels and selection rules for transitions, the quantum state of fluxonium can be protected in two ways. The first is to operate at the static or dynamical flux sweet spots~\cite{mundada2020floquet}, taking advantage of its insensitivity to dephasing from flux noise. The second is through an engineered decay-free subspace~\cite{earnest2018realization, lin2018demonstration}. More recently, coherence times above 1 millisecond were achieved on a fluxonium qubit with its primary transition energy a fraction of the thermal energy $k_B T$ at the flux insensitive spot~\cite{somoroff2021millisecond}. To fully exploit the potential of fluxnoium, it is therefore crucial to understand the loss mechanisms that affect and ultimately limit its performance.

Despite a wealth of work to this end, the dominant loss mechanism of fluxnoium remains elusive. Many attribute the dominant source to a phenomenological dielectric loss model~\cite{earnest2018realization, lin2018demonstration, nguyen2019high, zhang2021universal}, yet signatures of loss induced by out of equilibrium quasiparticles~\cite{pop2014coherent, vool2014non} were also observed. Other plausible sources include inductive loss~\cite{hazard2019nanowire} or two-level-system (TLS) defects~\cite{spiecker2022a} in some more exotic designs with highly disordered superconducting materials.

An underlying difficulty originates from the much larger parameter space, compared with that of transmon, that one needs to explore. Depending on the circuit parameters and the operation point, the qubit frequency as well as the transition matrix elements of both the charge and flux operators could vary by orders of magnitude, thus exposing the qubit to vastly different noise sources. The challenge for a systematic characterization of fluxonium's loss mechanisms is further compounded by the unavoidable fabrication and cryogenic-cycle variations, and the temporal fluctuations of coherence due to fluctuating TLS~\cite{klimov2018fluctuations} or quasiparticles~\cite{gustavsson2016suppressing}. Previous studies~\cite{earnest2018realization, lin2018demonstration, nguyen2019high, zhang2021universal} are limited in the range of parameters, largely focusing on decoherence at the sweet spot or its flux dependence on a narrow sets of circuit parameters.

\begin{figure}
    \centering
    \includegraphics[width=86mm]{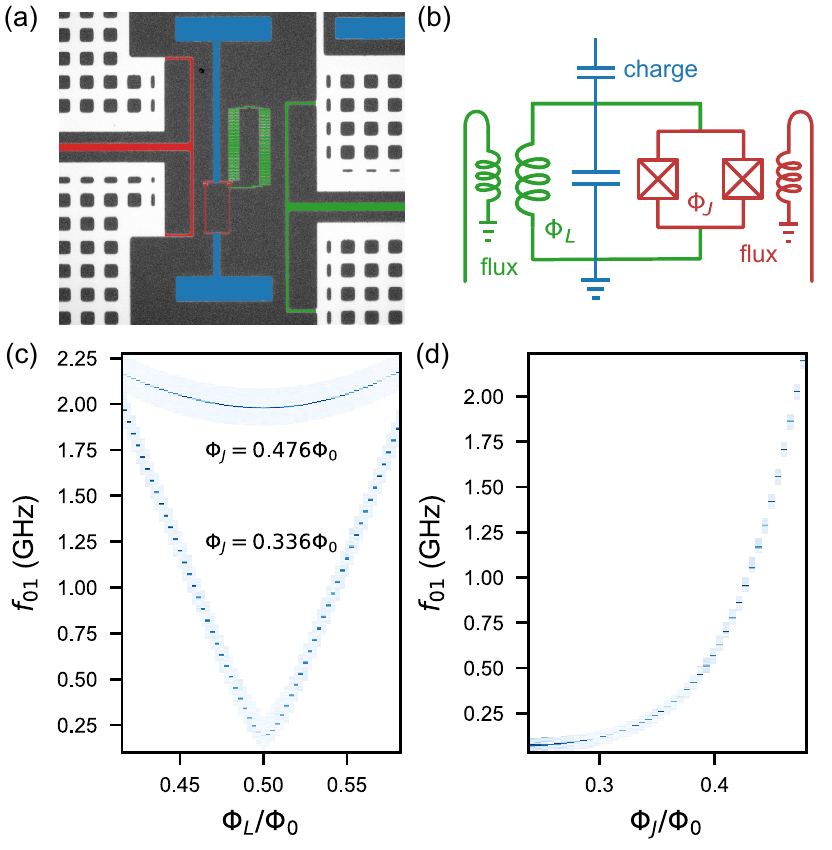}
    \caption{\label{fig:setup}(a) Optical image of an $E_J$-tunable fluxonium qubit made of aluminum (colored and white) on a silicon substrate (black). (b) Circuit model of the qubit. The circuit components are false-colored as in (a). The qubit pads as a shunt capacitor and the capacitively-coupled charge line for qubit excitation are marked in blue. The inductively-coupled flux lines and their threading loops are marked in green and dark red for $\Phi_{L}$ and $\Phi_J$, respectively. (c) Spectroscopy of the qubit as a function of $\Phi_{L}$ at two values of $\Phi_J$. (d) Spectroscopy of the $E_J$-tunable fluxonium as a function of $\Phi_J$ at $\phiext= \Phi_0/2$.}
\end{figure}

In this letter, we implement a fluxonium qubit with \insitu tunability on the Josephson energy $E_J$ in addition to the external flux bias. Compared with the original fluxonium, we replace the single weak junction with a DC superconducting quantum interference device (SQUID)~\cite{sete2017charge,lin2018demonstration}, thus the circuit is tunable by the flux threading the loop formed by the superinductor $\Phi_L$ and the flux threading the SQUID loop $\Phi_J$ provided by individual flux bias lines (\autoref{fig:setup}(a) and (b)). The circuit Hamiltonian is given by
\begin{equation}
    \hat{H}= 4 E_C \hat{n}^{2}+\frac{E_L}{2}\left(\hat{\varphi} + \frac{\Phi_{\text{ext}}}{\varphi_0}\right )^{2}-E_J \left(\Phi_J\right)\cos\hat{\varphi}\ , \label{eq:qubit_hamiltonian_effective}
\end{equation}
where $E_C$, $E_L$ and $E_J$ are the charging, inductive, and Josephson energies, respectively, $\hat \varphi$ and $\hat n$ are the superconducting phase across the SQUID and its conjugate variable, $\varphi_0 = \Phi_0/2\pi$ is the reduced flux quantum, and $\Phi_{\text{ext}}$ is the normalized external flux, with its value determined by $\Phi_L$ and $\Phi_J$. This Hamiltonian can be regarded as that of a fluxonium qubit with $E_J(\Phi_J)$ and $\Phi_{\text{ext}} = \Phi_L - \Phi_L^{(0)}(\Phi_J)$ independently tunable by two magnetic flux controls, where $\Phi_L^{(0)}(\Phi_J)$ is a small flux offset in the superinductor loop for the parameters used in the following experiments. The precise form of the dependencies is presented in the Supplementary Material. 

The device fabrication and measurement setup are similar to those presented in Ref.~\cite{bao2022fluxonium}. The qubit parameters $E_C/h=1.49~\GHz$, $E_L/h=0.65~\GHz$, $E_{J1}/h=7.12~\GHz$, and $E_{J2}/h=7.07~\GHz$ are obtained from a fit to the qubit spectra shown in \autoref{fig:setup}(c, d), where $E_{J1}$ and $E_{J2}$ are the Josephson energies of the two junctions forming the SQUID respectively. The qubit effective Josephson energy $E_J(\Phi_J)$ can be tuned between $0.05 - 14.19~\GHz \times h$. As a result, we are capable of \insitu tuning this single superconducting circuit from a heavy fluxonium~\cite{earnest2018realization}, through a traditional fluxonium qubit~\cite{manucharyan2009fluxonium}, to a weakly nonlinear resonator, by continuously adjusting $E_J$ from its maximum to minimum values.

\begin{figure}
    \includegraphics[width=86mm]{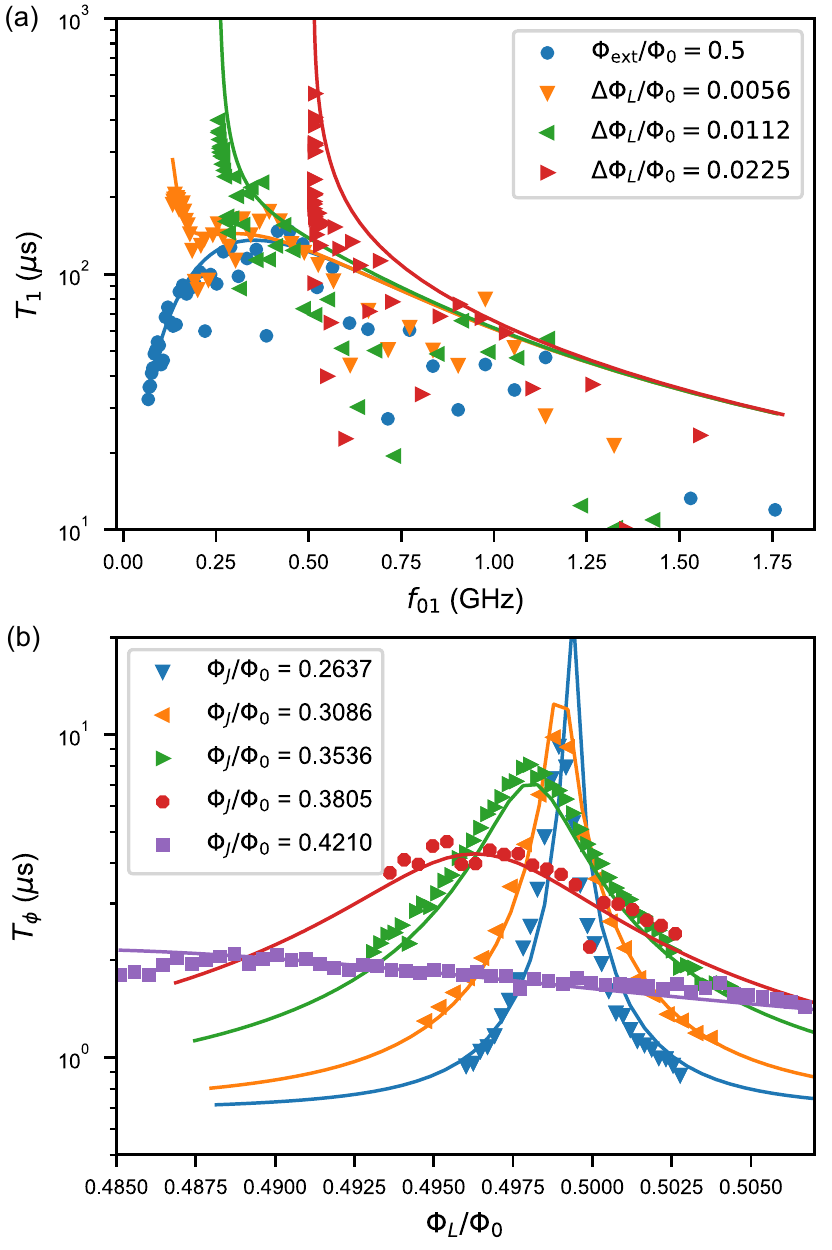}
    \caption{\label{fig:deco}(a) Qubit relaxation time $T_1$ versus the qubit frequency with the flux difference from the sweet spot $\Delta\Phi_L$ at a few fixed values while varying $E_J$. Solid lines represent simulated $T_1$ values using the combination of dielectric loss and loss due to $1/f$ flux noise. (b) Qubit dephasing time $T_{\phi}$ versus $\Phi_L$ around the sweet spot at several fixed values of $E_J$ defined by $\Phi_J$. Solid lines are fit to Gaussian dephasing function.}
\end{figure}

In \autoref{fig:deco}(a), we show the relaxation time $T_1$ versus the qubit frequency with $\phiext$ at a few fixed values close to the sweet spot by varying $E_J$. In contrast to the commonly used dielectric loss model~\cite{nguyen2019high, zhang2021universal}, where $T_1$ increases monotonically with increasing $E_J$ and decreasing qubit frequency, we observe that $T_1$ at the sweet spot ($\phiext = \Phi_0/2$, blue dots) peaks at a qubit frequency of around 400~MHz. This decreasing trend of $T_1$ with reduced qubit frequency resembles that of flux qubits~\cite{yan2016flux, quintana2017observation}, where it is attributed to $1/f$ flux noise with its $1/f$ form noise spectrum extending from very low frequencies to the GHz range. To account for this trend, therefore, we use a loss model $1/T_1 = \Gamma_1^{\text{diel}} + \Gamma_1^{1/f}$ with the combination of dielectric loss and loss due to $1/f$ flux noise in the following form:
\begin{align}
\Gamma_1^{\text{diel}} & = \frac{\hbar \omega_{01}^{2}}{4 E_C}  \abs{\mel{0}{\hat{\varphi}}{1}}^2 \tan\delta_C(\omega_{01}) \coth\left( \frac{\hbar \omega_{01}}{2k_B\Teff}\right)\ , \label{eq:t1_dielectric} \\
\Gamma_1^{1/f} & = \frac{E_L^2}{\hbar^2 \varphi_0^2}\abs{\mel{0}{\hat{\varphi}}{1}}^2S_{\Phi_L}(\omega_{01}) \left(1+ \exp\left(-\frac{\hbar\omega_{01}}{k_B\Teff}\right) \right)\ , \label{eq:t1_flux}
\end{align}
where the dielectric loss tangent and flux noise are written as $\tan\delta_C(\omega) = \tan\delta_C(\omega_r)\times\left(\omega/\omega_r\right)^{\epsilon}$ and $S_{\Phi_\beta}(\omega) = 2\pi A_\beta^2/\omega$ with $A_\beta$ being the $1/f$ flux noise amplitude in the external flux variable $\Phi_\beta$ with $\beta = L \text{ or } J$. We find that the flux noise in $\Phi_J$ has a negligible contribution to the qubit relaxation, thus is omitted from the model (see Supplementary Material). We extract $\tan\delta_C(\omega_r) = 2.0\times 10^{-6}$, $\epsilon = 0.2$, and $A_L = 14~\mu\Phi_0/\sqrt{\text{Hz}}$, where we assume $\Teff = 15$~mK from the equilibrium qubit population and $\omega_r = 2\pi \times 1$~GHz as a reference frequency. The simulated $T_1$ in the vicinity of the sweet spot (solid lines) shows a good agreement between the phenomenological model and the experimental result.

We additionally measure qubit dephasing versus the two flux controls. We find that a Gaussian decay function fits all the data well, which is consistent with dephasing due to $1/f$ noise~\cite{ithier2005decoherence, yoshihara2006decoherence}. In \autoref{fig:deco}(b), we show the Gaussian dephasing time $T_{\phi}$ obtained from spin-echo measurement versus $\phiext$ at different values of $E_J$. When $E_J$ is large, we find that $T_{\phi}$ peaks at $\phiext \approx \Phi_0/2$ corresponding to the flux insensitive point with respect to $\phiext$. However, as $E_J$ decreases, the maximum $T_\phi$ point with respect to $\phiext$ shifts further away from $\Phi_0/2$. This agrees well with a model with correlated $1/f$ flux noise in both loops of the qubit~\cite{gustavsson2011noise}. Fitting all the data, we find $A_L=12~\mu\Phi_0/\sqrt{\text{Hz}}$, $A_J=7.6~\mu\Phi_0/\sqrt{\text{Hz}}$, and a correlation factor $c_{LJ}=0.51$ (see Supplementary Material). The flux noise amplitude $A_L$ here is close to that obtained from the relaxation measurement. In fact, if we use a $1/f$ flux noise model $S_{\Phi_L}(\omega) = 2\pi A_L^2/\omega^\alpha$ with a variable exponent $\alpha$, $A_L=12~\mu\Phi_0/\sqrt{\text{Hz}}$ at 1~Hz and $\alpha=0.99$ can describe both $T_1$ and $T_{\phi}$ data simultaneously.

We note that the dielectric loss we extracted is close to some of the most coherent fluxonium qubits that have been demonstrated~\cite{nguyen2019high, somoroff2021millisecond}, however our coherence at low frequencies at the sweet spot is limited by $1/f$ flux noise with a much higher amplitude. This flux noise level could be related to the geometry of the Josephson junction array and superconducting wires forming the qubit loops~\cite{lanting2009geometrical, anton2013magnetic, braumuller2020characterizing}, thus can be further optimized for higher coherence.

\begin{figure}
    \includegraphics[width=86mm]{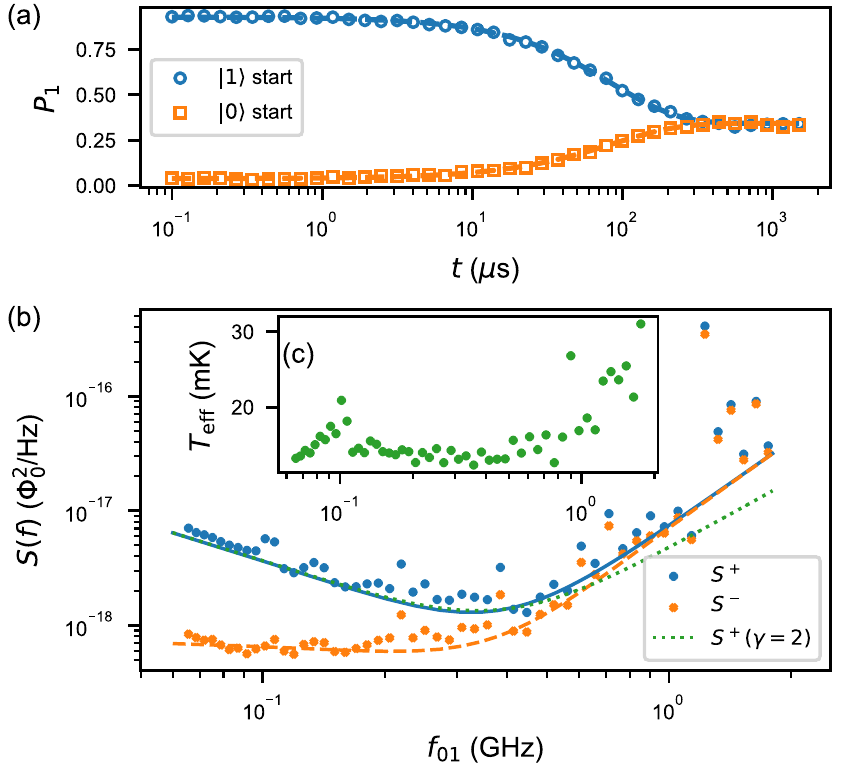}
    \caption{\label{fig:noise_density}(a) Representative experimental data of qubit relaxation from the excited (blue circle) and ground (orange square) state. (b) The extracted symmetric $S^+(\omega)$ and anti-symmetric $S^-(\omega)$ noise power spectra versus the qubit frequency in terms of effective flux noise. The solid lines are fit to a model with a combination of flux noise and dielectric loss. The inset figure shows the effective temperature versus qubit frequency estimated from the measured relaxation rate.}
\end{figure}

To further understand the loss mechanisms beyond the phenomenological model, we perform noise spectroscopy in a frequency range around $k_B T/h$ at $\phiext = \Phi_0/2$ where the qubit relaxation is the most sensitive to the relevant noises because the transition matrix element $\abs{\mel{0}{\hat{\varphi}}{1}}$ is at its maximum. With high-fidelity qubit reset (see Supplementary Material), we are able to map out the noise spectra $S(\omega)$ and $S(-\omega)$ at the positive and negative frequencies by measuring the relaxation rate $\Gamma_{\downarrow}$ and $\Gamma_{\uparrow}$ from the excited and ground state, respectively. The representative relaxation data is shown in \autoref{fig:noise_density}(a). 

In \autoref{fig:noise_density}(b), we plot the extracted noise spectra in terms of symmetric $S^+(\omega) = S(\omega) + S(-\omega)$ and anti-symmetric $S^-(\omega) = S(\omega) - S(-\omega)$ as effective flux noise, where the noise power spectral density is directly obtained from the relaxation rate by $\Gamma_{\downarrow /\uparrow} = (E_L/\hbar/\varphi_0)^2 \abs{\mel{0}{\hat\varphi}{1}}^2 S(\pm\omega)$. The effective temperature is computed as $\exp(-\hbar\omega/(k_B \Teff)) = \Gamma_\uparrow/\Gamma_\downarrow$, as shown in the inset figure in \autoref{fig:noise_density}(b). We fit the data with a model given by $S^+(\omega)=2\pi A_L^2/\omega^{\alpha}(1+\exp(-\hbar\omega/k_BT_A))+ (\hbar^3\varphi_0^2/(4E_C E_L^2)) \tan\delta_C\times \omega^{\gamma}$ and $S^-(\omega) = S^+(\omega) \tanh(\hbar\omega_{01}/2k_B T_A)$. We find that $\alpha\approx 1$, $\gamma\approx2.5$, and $T_A = 13$~mK can describe the data well. In fact, the measurement and data analysis protocols for the noise spectroscopy are very similar to what was demonstrated in a flux qubit~\cite{quintana2017observation} for the quantum annealing purpose~\cite{johnson2011quantum}. Due to the reduced sensitivity to flux noise in fluxonium, we are able to achieve two-orders of magnitude longer coherence times and unveil the noise spectra originated from difference sources. 

Same as in the $T_1$ data, a clear turning point in $S^+(\omega)$ at near 400~MHz signifies the transition from dielectric loss to flux noise. The frequency of this turning point coincides with the frequency $\omega \approx 2k_B \Teff /\hbar$ around which the two-sided noise spectrum undergoes a transition from quantum noise ($S^+(\omega)\approx S^-(\omega)$) to classical noise ($S^+(\omega) \gg S^-(\omega)$). 

Below the transition frequency, $S^+(\omega)$ roughly follows the $1/f^\alpha$ form and $S^-(\omega)$ asymptotically approaches to a constant when frequency decreases below 100~MHz. We also observe a nearly constant $\Teff\approx 15~\text{mK}$ in this regime which indicates that the system is stabilized in a relatively uniform and cold environment.

Above the transition frequency, a fit to the lower bound of the noise spectra approximately gives $S^+(\omega)\approx S^-(\omega) \propto \omega^{2.5}$. We find that the noise amplitude or relaxation rate can vary drastically by more than a factor of 3 from one frequency to another in a reproducible way, contributing a large uncertainty in the extracted functional form of the noise spectrum. Moreover, at frequencies $>1$~GHz, the population decay sometimes deviates from the exponential decay model. These behaviors are commonly understood as resonant interaction with TLS distributed randomly in the relevant range of qubit frequency in superconducting circuits~\cite{martinis2005decoherence, barends2013coherent, muller2019towards}. Apart from the enhanced relaxation at the frequencies coincide with a TLS, the background relaxation, usually attributed to dielectric loss, can be derived from the perturbation theory assuming weak electric-dipole interaction with a bath of TLS uniformly distributed in frequency~\cite{martinis2005decoherence}. Considering the TLS bath in thermal equilibrium, the relaxation due to TLS loss can be written as $\Gamma_1 = \Gamma_\uparrow + \Gamma_\downarrow = (\hbar \omega_{01}^{2}/4 E_C)  \abs{\mel{0}{\hat{\varphi}}{1}}^2 \tan\delta_C$ therefore the noise spectrum should have a form $S^+(\omega)\propto \omega^2$ (see Supplementary Material and Ref.~\cite{shnirman2005low}). We note that this loss derived from the TLS theory is different from the phenomenological dielectric loss model in \autoref{eq:t1_dielectric} as $S^+(\omega)$ given by TLS loss is temperature independent. The noise spectra given by a dense weakly-coupled TLS bath constitutes a background noise for relaxation, shown as the green dashed lines in \autoref{fig:noise_density}(b).

\begin{figure*}
\includegraphics[width=1.0\textwidth]{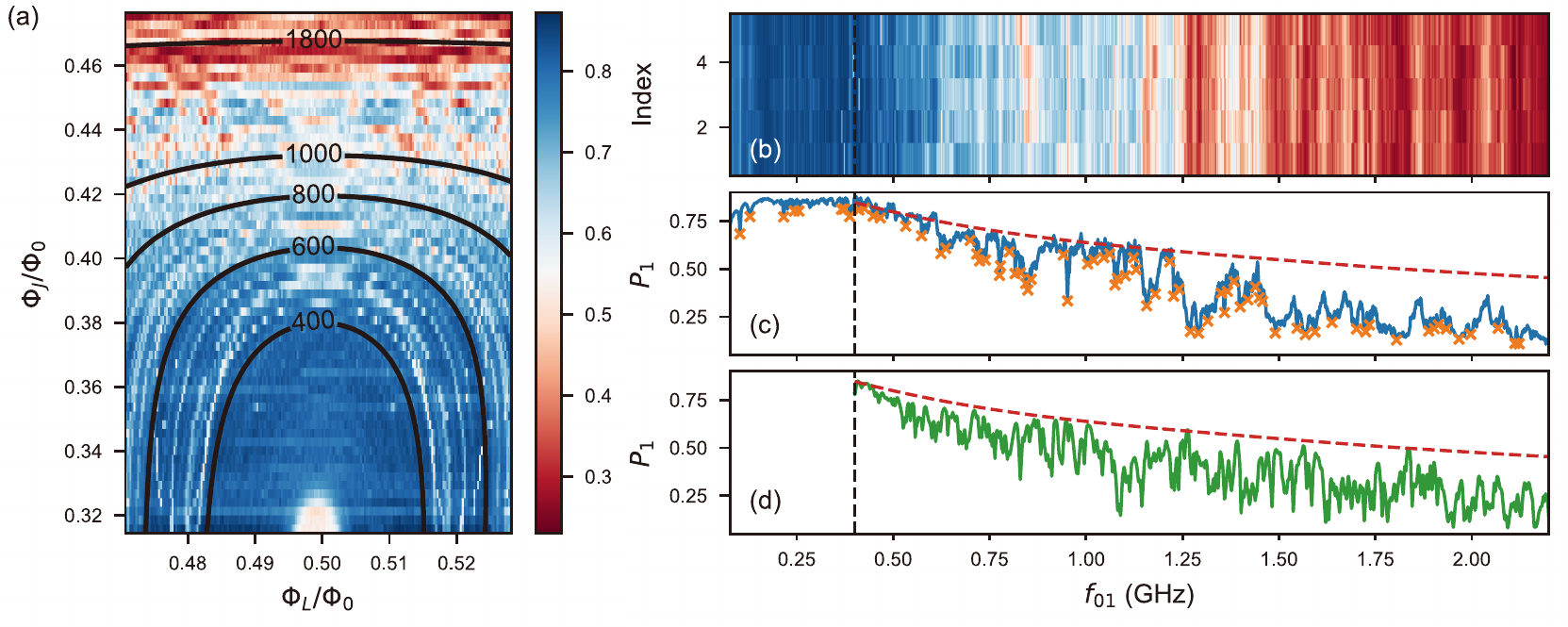}
\caption{\label{fig:tls}(a) Fixed delay ($\tau= 15~\mu $s) $P_1$ map in the $\phiext$ range $\Phi_0/2\pm 0.03 \Phi_0$ around its sweet spot and the $\Phi_J$ range from $0.32\Phi_0$ to $0.47\Phi_0$. The black solid lines mark the qubit frequency $f_{01}$ contours (in $\MHz$) as a function of $\Phi_L$ and $\Phi_J$. (b) Repeated measurements of $P_1$ versus the qubit frequency at the same fixed delay. The sweeps are performed in an equal $0.5~\MHz$ frequency step. (c) Averaged $P_1$ data with $P_1$-dips labeled by crosses. The prominence of the peak is set to 0.02 for peaking finding. (d) Simulated $P_1$ curves corresponding to the experimental conditions. The vertical black dashed line through (b), (c) and (d) marks the loss mechanism turning frequency at $400~\MHz$. The red dashed lines in (c) and (d) are the $P_1$ analytically derived from a dense and weakly coupled TLS bath.}
\end{figure*}

We next perform swap spectroscopy~\cite{mariantoni2011photon} along two flux axes $\Phi_J$ and $\phiext$ with a fine step for a detail study of the energy relaxation in the dielectric or TLS loss limited regime. To improve measurement speed, we choose to measure the excited state population after a specific wait time $\tau$ in the qubit decay process following a qubit initialization in the excited state. As shown in \autoref{fig:tls}(a), we acquire the $P_1$ at $\tau= 15~\mu $s, covering a qubit frequency range from $160~\MHz$ to 2~GHz. Comparing with the qubit frequency $f_{01}$ contours as shown by the solid lines in \autoref{fig:tls}(a), we find that the $P_1$-dip contours closely follow the qubit frequency contours. This strongly suggests that the positions of these $P_1$-dips are defined by a set of fixed qubit frequencies. This behavior agrees well with the typical TLS model, where each contour indicates the fixed frequency of a TLS that causes enhanced relaxation when the qubit is tuned in resonant with the TLS. We note that we also observe some fluctuating TLS in frequency, which is consistent with previous observations in transmon qubits~\cite{klimov2018fluctuations, carroll2021dynamics}. 

To acquire quantitative information on the TLS bath, we repeatedly measure the qubit $P_1$ at $\tau =15~\mu$s while sweeping the qubit frequency by tuning $E_J$ and maintaining $\phiext = \Phi_0/2$. As shown in \autoref{fig:tls}(b), the $P_1$ measurement is repeated 5 times at each frequency and the whole experiment takes 40 hours. There is an obvious trend that $P_1$ decreases as qubit frequency increases, which is consistent with the background noise spectrum (dashed red line) of dielectric loss $S^+(\omega) \propto \omega^2$. The extracted dips with respect to the background in the $P_1$ curve are labeled as TLS as shown in \autoref{fig:tls}(c).

We find no clear trend of TLS density on frequency but the dip depth and impacted bandwidth of each TLS decreases with decreasing frequency. As a result, the available TLS-free frequency region is larger for the fluxonium at lower qubit frequency by increasing $E_J$. This can be attributed to a decreasing transition element of the charge operator $\abs{\mel{0}{\hat n}{1}}$ with decreasing qubit frequency, \textit{i.e.}, the electric-dipole interaction between the qubit and a TLS is suppressed at low qubit frequency for fluxonium. We confirm this hypothesis by performing numerical simulations and quantitatively obtaining the same behavior as shown in \autoref{fig:tls}(d) (see Supplementary Material). 

In summary, we implement a fluxonium qubit with \insitu tunability of its Josephson energy for the study of the underlying decoherence mechanisms. We measure the noise spectrum responsible for the qubit relaxation from approximately 80~MHz to 2~GHz, covering well the frequency relevant to the thermal fluctuation energy $k_B\Teff /h \approx 300~\MHz$. In the large explored $E_J$ parameter regime around the $\phiext = \Phi_0/2$ flux sweet spot, the relaxation can be explained by loss due to electric-dipole interaction with TLS defects and $1/f$ flux noise, with a transition point of these two noise spectra at around 400~MHz.

Above the transition frequency, we observe a $f^\gamma$ power-law noise spectrum, with $\gamma \gtrsim 2$, originated from charge TLS, i.e., the electric-dipole interaction with TLS. We show strong evidences that these TLS impose a stringent limit on the coherence times and the stability of fluxonium as the case for transmon. However, due to the approximate $f^2$ frequency dependence of the TLS-induced relaxation and the low qubit frequency with large anharmonicity by design, fluxonium can couple less to these TLS with lowered qubit frequencies before hitting the $1/f$ flux noise limit. On the contrary, weakly anharmonic qubits like transmon only has a linear frequency dependence on the relaxation due to TLS, and are constrained at relatively higher qubit frequencies ($\gtrsim 4$~GHz) desired by qubit operations. 

Our finding demonstrates a key advantage of fluxonium in qubit coherence: decoupling to charge TLS at low qubit frequencies. This advantage could pave the way for fluxonium to become the mainstream qubit of choice for the implementation of high-fidelity superconducting quantum processors.

\begin{acknowledgments}
We thank all full-time associates at Alibaba Quantum Laboratory for experimental support and Xin Wan for insightful discussions.
\end{acknowledgments}

\bibliography{ref}

\end{document}


\title{Supplementary Material for \\``Characterization of loss mechanisms in a fluxonium qubit''}

\author{Hantao Sun}
\thanks{These two authors contributed equally}
\affiliation{Alibaba Quantum Laboratory, Alibaba Group, Hangzhou, Zhejiang 311121, China}
\author{Feng Wu}
\thanks{These two authors contributed equally}
\affiliation{Alibaba Quantum Laboratory, Alibaba Group, Hangzhou, Zhejiang 311121, China}
\author{Hsiang-Sheng Ku}
\affiliation{Alibaba Quantum Laboratory, Alibaba Group, Hangzhou, Zhejiang 311121, China}
\author{Xizheng Ma}
\affiliation{Alibaba Quantum Laboratory, Alibaba Group, Hangzhou, Zhejiang 311121, China}
\author{Jin Qin}
\affiliation{Alibaba Quantum Laboratory, Alibaba Group, Hangzhou, Zhejiang 311121, China}
\author{Zhijun Song}
\affiliation{Alibaba Quantum Laboratory, Alibaba Group, Hangzhou, Zhejiang 311121, China}
\author{Tenghui Wang}
\affiliation{Alibaba Quantum Laboratory, Alibaba Group, Hangzhou, Zhejiang 311121, China}
\author{Gengyan Zhang}
\affiliation{Alibaba Quantum Laboratory, Alibaba Group, Hangzhou, Zhejiang 311121, China}
\author{Jingwei Zhou}
\affiliation{Alibaba Quantum Laboratory, Alibaba Group, Hangzhou, Zhejiang 311121, China}
\author{Yaoyun Shi}
\affiliation{Alibaba Quantum Laboratory, Alibaba Group USA, Bellevue, WA 98004, USA}
\author{Hui-Hai Zhao}
\email{huihai.zhh@alibaba-inc.com}
\affiliation{Alibaba Quantum Laboratory, Alibaba Group, Beijing 100102, China}
\author{Chunqing Deng}
\email{chunqing.cd@alibaba-inc.com}
\affiliation{Alibaba Quantum Laboratory, Alibaba Group, Hangzhou, Zhejiang 311121, China}

\maketitle

\bookmarksetup{startatroot}

\section{$E_J$-tunable Fluxonium Hamiltonian}
The Hamiltonian of the $E_J$-tunable fluxonium can be written as
\begin{equation}
    \hat{H} = 4E_C\hat{n}^2+\frac{1}{2}E_L\hat{\varphi}^2 - E_{J1}\cos(\hat{\varphi} -\varphi_L+\varphi_J/2) -E_{J2}\cos(\hat{\varphi}-\varphi_L-\varphi_J/2), \label{eq:ej_tunable_hamiltonian_original}
\end{equation}
where $\varphi_L = \Phi_L/\varphi_0$ is the effective external reduced magnetic flux on the qubit loop and $\varphi_J = \Phi_J/\varphi_0$ is that in the SQUID loop.

Considering the design of two-loop structure, the effective flux and the physical flux has the relationship due to the qubit geometry~\cite{orlando1999superconducting,gustavsson2011noise}
\begin{align}
    \varphi_L & = \varphi_1 + \varphi_2 / 2, \nonumber \\
    \varphi_J & = \varphi_2, \label{eq:phij_to_phi12}
\end{align}
where $\varphi_1$ is the external reduced magnetic flux applied in the loop defined by the junction array and $\varphi_2$ is that applied in the SQUID loop. From these equations, the effective reduced magnetic flux has a native crosstalk of 0.5 from $\varphi_J$ to $\varphi_L$.

To compare with the conventional fluxonium Hamiltonian, the potential term in \autoref{eq:ej_tunable_hamiltonian_original} can be rewritten as
\begin{align}
    E_{J1}& \cos(\hat{\varphi}-\varphi_L+\varphi_J/2)+E_{J2}\cos(\hat{\varphi}-\varphi_L-\varphi_J/2) \nonumber                                       \\
    = & \mathrm{sgn}\left[\cos\left(\frac{\varphi_J}{2}\right)\right]\sqrt{E_{J1}^2+E_{J2}^2+2E_{J1}E_{J2}\cos\varphi_J} \cos\left(\hat{\varphi}-\varphi_L+\arctan\left(\frac{E_{J1}-E_{J2}}{E_{J1}+E_{J2}}\tan\frac{\varphi_J}{2}\right)\right)\nonumber \\
    = & E_J(\varphi_J)\cos\left(\hat{\varphi}-\varphi_\text{ext}(\varphi_J,\varphi_L)\right).
    \label{eq:ej_tunable_potential_effective}
\end{align}
Now we can see the $E_J$-tunable fluxonium at any $(\varphi_J,\varphi_L)$ can be treated as a conventional fluxonium Hamiltonian with effective $E_J(\varphi_J)=\mathrm{sign}\left[\cos(\varphi_J/2)\right]\sqrt{E_{J1}^2+E_{J2}^2+2E_{J1}E_{J2}\cos\varphi_J}$ and effective $\Phiext/\varphi_0 = \varphi_\text{ext}(\varphi_J,\varphi_L)=\varphi_L- \varphi_L^{(0)}$ where $\Phi_L^{(0)}/\varphi_0 = \varphi_L^{(0)} = \arctan\left(\frac{E_{J1}-E_{J2}}{E_{J1}+E_{J2}}\tan\frac{\varphi_J}{2}\right)$ is a $\varphi_J$-dependent flux offset in the qubit loop. Note there is a $\mathrm{sign}$ function in $E_J(\varphi_J)$, and a $\tan$ function in $\varphi_\text{ext}(\varphi_J,\varphi_L)$, both of which are singular at $\varphi_J=\pi/2+2\pi n~(n\in \mathbb{Z})$. These will give a sudden change of the flux sweet spot in $\Phiext$ near the singular point, which is verified in our experiment.

\section{Flux crosstalk}
\begin{figure}
    \includegraphics[width=0.5\textwidth]{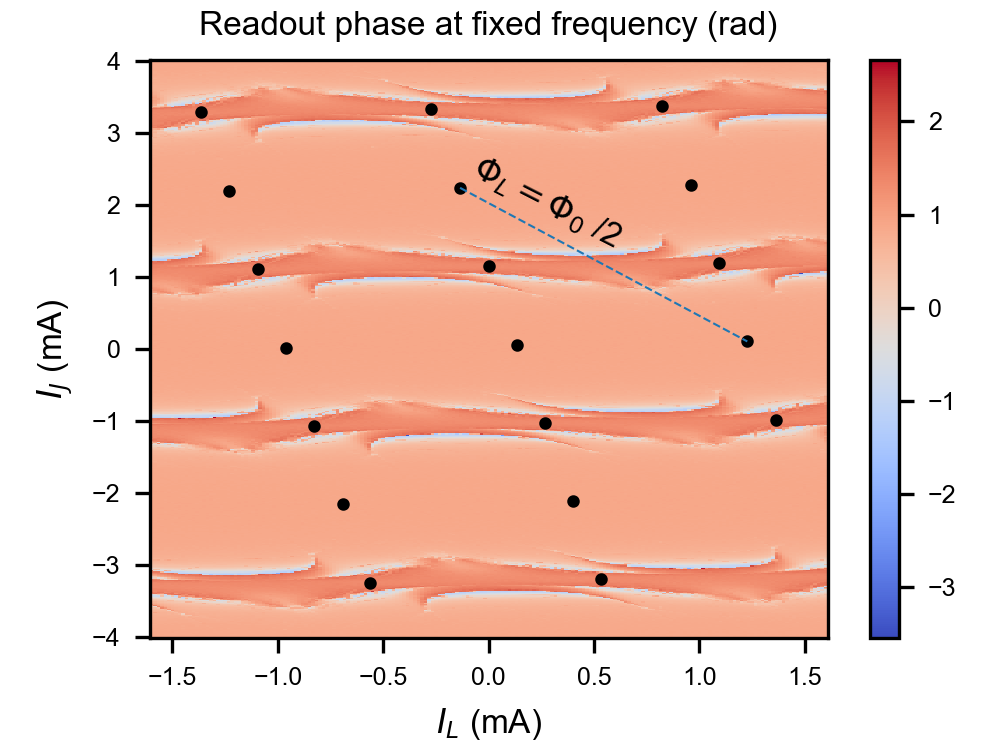}
    \caption{\label{fig:cavity_responds}Fixed frequency readout transmission phase versus the bias currents for $\Phi_J$ and $\Phi_L$. The readout frequency is chosen near the bare resonance frequency of the readout resonator. The black dots are symmetry points where $\Phi_L$ and $\Phi_J$ are integer multiples of $\Phi_0/2$. The blue dashed line represents bias points with $\Phi_L = \Phi_0/2$ in one period of $\Phi_J$ varying from 0 to $\Phi_0/2$.}
\end{figure}

\begin{figure}
    \includegraphics[width=0.8\textwidth]{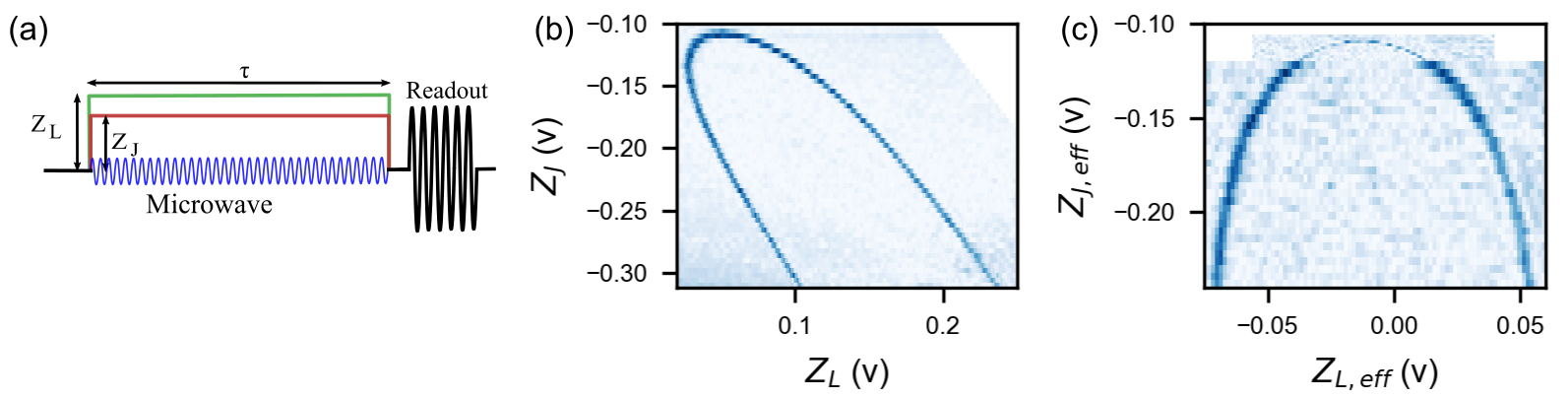}
    \caption{\label{fig:freq_contour}(a) Pulse sequence used to measure the qubit frequency contours with different amplitude of the two flux biases. (b) Qubit frequency contour with 500~$\MHz$ driving before crosstalk correction. (c) Qubit frequency contour with 500~$\MHz$ driving after crosstalk correction.}
\end{figure}

We search for operation points and calibrate the flux crosstalk by first scanning the readout resonator's response to the two flux biases. Using standard S-parameter measurement, the fixed frequency readout transmission phase was measured near the resonator bare frequency, and its dependence on the two flux biases is shown in \autoref{fig:cavity_responds}. The black dots in the figure mark the symmetry points where $\Phi_L$ and $\Phi_J$ are integer multiples of $\Phi_0/2$.
The dashed blue line represents points with $\Phi_L = \Phi_0/2$ in one period of $\Phi_J$. The $\Phiext=\Phi_0/2$ points (the sweet spots) are always in the adjacent regions along the line. Using two static biases, the qubit was biased at $\Phiext=\Phi_0/2$ points as the coarse operation points of measurements.

As discussed in the previous section, the two control flux biases $\Phi_J$ and $\Phi_L$ have an intrinsic crosstalk originated from \autoref{eq:phij_to_phi12}. An additional magnetic field crosstalk also exists between the two biases. After finding coarse operation points, we implement a qubit frequency contour measurement to calibrate the crosstalk. The pulse sequence is shown in \autoref{fig:freq_contour}(a). A microwave excitation with a duration $\tau=50~\mu$s at a fixed frequency is applied to the qubit followed by qubit readout. During the qubit drive, flux pulses with amplitudes $Z_J$ and $Z_L$ are applied to shift the qubit frequency. A typical qubit frequency contour with 500~MHz driving is shown in \autoref{fig:freq_contour}(b). Obviously, the frequency contour is asymmetric due to the crosstalk between $\Phi_J$ and $\Phi_L$. Based on measured frequency contours at several driving frequencies, we extracted a linear crosstalk correction matrix to achieve orthogonal flux control:
\begin{equation}
\begin{bmatrix} Z_{L}\\ Z_{J} \end{bmatrix} = \begin{bmatrix} 1 & 0.57621 \\ 0.02358 & 1 \end{bmatrix} ^{-1} \cdot \begin{bmatrix} Z_{L, \text{eff}}\\ Z_{J, \text{eff}} \end{bmatrix}.
\end{equation}
This simple correction matrix works well in the range of the experiments when $\Phi_J$ is away from its $\Phi_0/2$ point. The $500~\MHz$ frequency contour after crosstalk correction is shown in \autoref{fig:freq_contour}(c).

\section{Pulse measurement protocols}
The experiment is performed in a dilution refrigerator at a base temperature below 10~mK. The wiring setup inside the dilution refrigerator and electronic readout and control circuitries are similar to that described in a previous report~\cite{bao2022fluxonium}. Except here, 0-dB attenuation (a total of 30~dB in the cryostat) together with a 45~$\MHz$ low-pass-filter on mixing-stage are used for the flux control lines in adaptation with low qubit-frequencies and a large flux tuning range. We nominally use flux pulses with 10~ns rise/fall time to be within the bandwidth of the flux lines. 

Qubit state readout is performed via the dispersive interaction between the qubit and a resonator. A near-resonance microwave with 1.5~$\mu$s duration is applied to the readout cavity for homodyne detection. The qubit readout is performed at $\Phi_J=0.376\Phi_0$ and $\Phiext=\Phi_0/2$ ($\Phi_L=0.4995\Phi_0$). The qubit excitation pulses are calibrated at this bias point with a duration of about 30~ns. Qubit coherence times at all the other operation points are measured by actuating the flux pulses in the two flux controls as shown in \autoref{fig:t1_vs_Phi_L}(a).

\begin{figure}
    \includegraphics[width=0.5\textwidth]{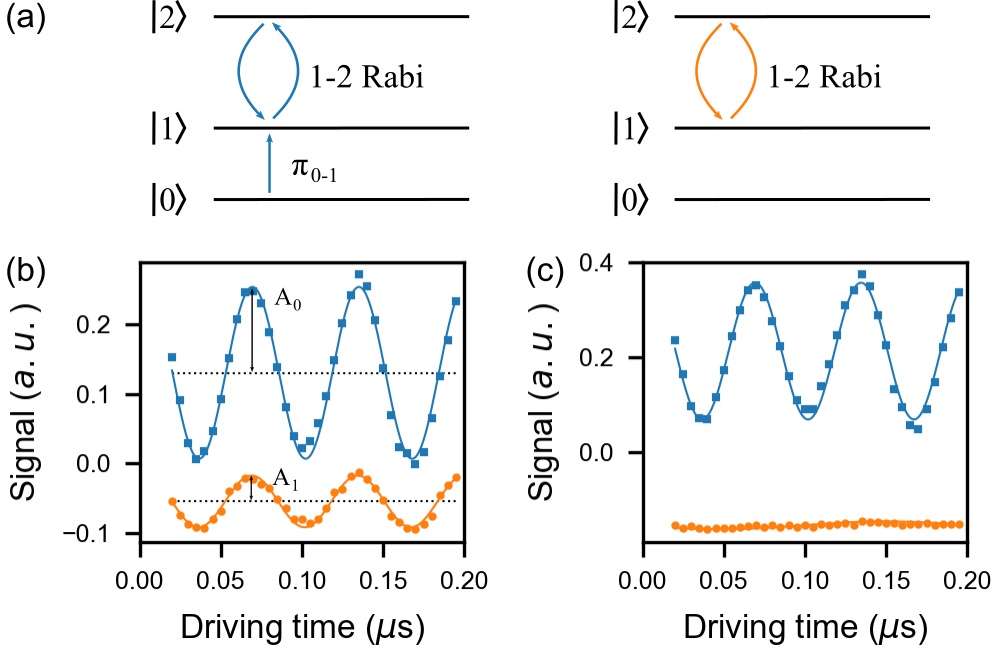}
    \caption{\label{fig:ef_rabi}(a) The protocol of the 1-2 Rabi experiment to evaluate the stray population of the qubit excited state. The qubit is prepared in the initial state (orange) or with its excited state population inverted by a $\pi_{01}$ pulse (blue). (b) The observed 1-2 Rabi oscillations with the qubit initialized in the thermal state. Fitting the two oscillations with sinusoidal functions yields signal amplitudes $A_0$ and $A_1$ that are proportional to the qubit population of the ground and excited states. (c) The observed 1-2 Rabi oscillations with active qubit initialization.}
\end{figure}

At the bias point for readout, the qubit transition frequency $f_{01}=385~\MHz$, which is close to the frequency of the thermal energy $k_B T/\hbar$. So at thermal equilibrium, the qubit always stabilizes at a mixed state. We perform reset on the qubit to better initialize qubit to its ground state. The reset scheme is similar to that described in Ref.~\cite{bao2022fluxonium} except that two fast-flux pulses are needed to offset the qubit off the $\Phiext= \Phi_0/2$ point for lifting the selection rule. Simultaneously, a 20~$\mu$s microwave pulse is applied to pump the red-sideband transition, to transfer the qubit excitation to the rapid damping 1-photon state of the resonator. To evaluate the stray population of the qubit excited state, we perform a 1-2 Rabi experiment~\cite{jin2015thermal}. The experimental protocol is shown in \autoref{fig:ef_rabi}(a), the qubit is prepared in its initial state or with its population inverted by a $\pi_{01}$-pulse to translate its excited/ground state population into the 1-2 Rabi signal amplitude. The observed 1-2 Rabi oscillations are shown in \autoref{fig:ef_rabi}(b) and \autoref{fig:ef_rabi}(c). Fitting to a sinusoidal function, the ground state signal amplitude $A_0$ and excited state signal amplitude $A_1$ are acquired. Then, the excited state stray population can be estimated as $P_1 = A_1/(A_1+A_0)$. With thermal excitation, as is shown in \autoref{fig:ef_rabi}(b), the stray population of the excited state is extracted as $P_1=23.14\%$, corresponding to an effective temperature $\Teff\approx 15.4~\text{mK}$. When qubit is initialized in ground state with the sideband reset, the 1-2 Rabi oscillations are shown in \autoref{fig:ef_rabi}(c). The stray population of the excited state after reset is extracted as $P_{1}=3.40\%$. After reset, the readout contrast is 89\%.

\section{Qubit parameters}
\begin{figure}
    \includegraphics[width=0.8\textwidth]{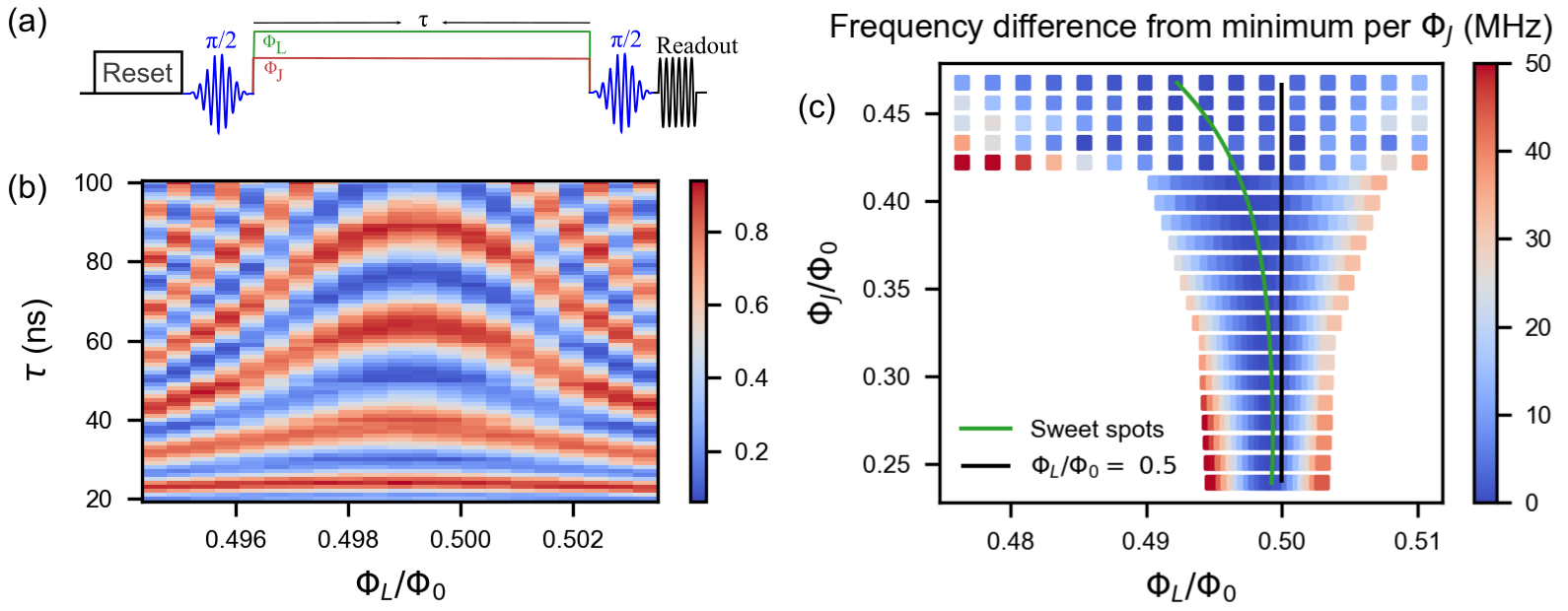}
    \caption{\label{fig:adjust_sweet_line}(a) Pulse sequence used to adjust the $\Phiext = \Phi_0/2$ ``sweet line'' using a Ramsey experiment with flux pulses. After the qubit is reset to its grounded state, the Ramsey interference experiment is followed with a 100~ns delay. During the duration $\tau$ of the Ramsey interference experiment, the qubit is tuned to the target $E_J$ and flux bias values using $\Phi_J$ and $\Phi_L$. (b) Typical Ramsey data at $\Phi_J=0.3418\Phi_0$ with different $\Phi_L$ in the small range around the $\Phiext = \Phi_0/2$ sweet spot. (c) With fast Fourier transform (FFT), the frequency difference from the minimum points with respect to $\Phi_L$ are extracted. The $\Phi_L$ of the minimum frequency point for each $\Phi_J$ is plotted as the green line correspond to the precise sweet spot where $\Phiext = \Phi_0/2$. The black line is the $\Phi_L/\Phi_0=0.5$ line for comparison.}
\end{figure}

With crosstalk correction, we managed to tune the qubit in the orthogonal flux bases. However, in the experiments we found that the sweet spot for each $\Phi_J$ is not exactly at $\Phi_L/\Phi_0=0.5$, especially when $\Phi_J$ is near to $\Phi_0/2$. This can be attributed to the intrinsic dependency of $\Phiext$ on $\Phi_J$, as shown in \autoref{eq:ej_tunable_potential_effective}. Experimentally, a correction term can be adapted to further fine-tuning the $\Phiext = \Phi_0/2$ sweet spots for each $\Phi_J$. We implemented a Ramsey experiment with flux pulses to find out the correction along $\Phi_J$. The pulse sequence for the Ramsey experiment is shown in ~\autoref{fig:adjust_sweet_line}(a). After the qubit initialized to its ground state with reset, a $\pi/2$-pulse is applied to bring the qubit to the equator on the Bloch sphere, followed by a delay of variable duration $\tau$, during which the qubit precesses alone its Z-axis. During the phase accumulating precession, the qubit is flux pulsed using different $\Phi_L$ and $\Phi_J$. Typical Ramsey oscillations versus $\Phi_L$ around the sweet spot with $\Phi_J/\Phi_0=0.3418$ is shown in \autoref{fig:adjust_sweet_line}(b). The qubit frequency as a function of $\Phi_J$ and $\Phi_L$ can be extracted using fast Fourier transform (FFT) on the Ramsey data. In \autoref{fig:adjust_sweet_line}(c), we plot the qubit frequency difference compared with the minimum positions of $\Phi_L$ for each $\Phi_J$. The minimum frequency point for each $\Phi_J$ as plotted by the green line corresponds to the precise sweet spot per $\Phi_J$, which is the $\Phiext/\Phi_0=0.5$ spot. We also plot the $\Phi_L/\Phi_0=0.5$ line in black for a comparison. The sweet spot ($\Phiext/\Phi_0=0.5$) is quite near with $\Phi_L/\Phi_0=0.5$ when $\Phi_J$ is far from $\Phi_0/2$, but deviate fast when $\Phi_J$ getting close to $\Phi_0/2$.

In the main text, we measure qubit spectroscopy in Fig.~1. Fitting the spectra to the full circuit model, we manage to extract the qubit parameters. However, we find large errors exist in the fitting. We attribute errors come from the two aspects: On one hand, full period qubit spectroscopy is hard to be measured because of the extra low qubit frequency toward $\Phi_J=0$. Note that the qubit is submerged in thermal flux noise when qubit frequency is in the pass-band of the 45~$\MHz$ low-pass-filter in our flux-control lines. On the other hand, possible errors in the crosstalk matrix and voltage-to-flux transformation coefficients can be a result of coarse spectroscopy. In the end, we find the Ramsey experiment with flux-pulses come up with abundant qubit spectroscopy data with high accuracy. This makes the combined fitting possible that includes the full circuit model, crosstalk matrix and voltage-to-flux transformation coefficients. We use a relation that can fully define the mapping from the applied flux-pulse voltages ($Z_{L,J}$) to magnetic fluxes ($\Phi_{L,J}$).
\begin{equation}
\begin{bmatrix} \Phi_{L}\\ \Phi_{J} \end{bmatrix} = \left(\begin{bmatrix} 1 & O_1 \\ O_2 & 1 \end{bmatrix} \cdot \begin{bmatrix} Z_{L}\\ Z_{J} \end{bmatrix} - \begin{bmatrix} Z_{0,L}\\ Z_{0,J} \end{bmatrix} \right) \cdot \begin{bmatrix} 1/S_{L}\\ 1/S_{J} \end{bmatrix}.
\end{equation}
Here, six parameters are used: the zero-point of the system after crosstalk correction $Z_{0, L}$ and $Z_{0,J}$; the scaling coefficients from voltage to flux $S_L$ and $S_J$; and the crosstalk matrix off diagonal matrix elements $O_1$ and $O_2$. Combining this relation with the full circuit model, given the initial guess from crosstalk calibration and qubit spectroscopy fitting, we fit the above the qubit frequency from Ramsey and acquire the accurate qubit parameters shown in the main text. The fitting mean absolute deviation (MAD) is about 4~$\MHz$. 

\section{Pulse distortion compensation}
\begin{figure}
    \includegraphics[width=0.5\textwidth]{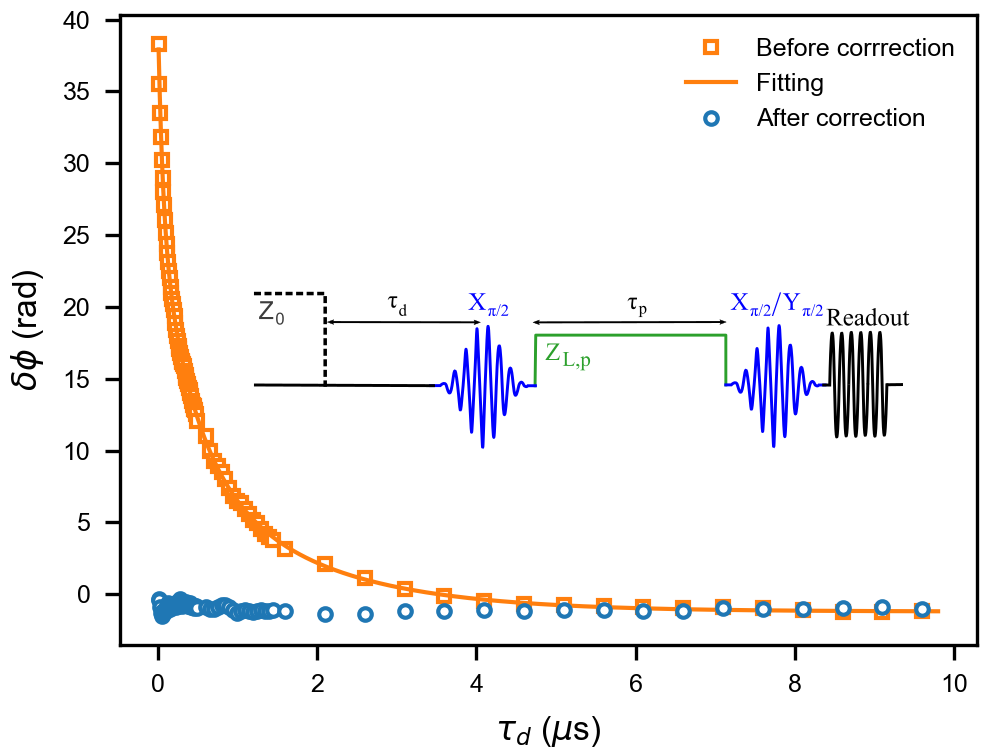}
    \caption{\label{fig:Zpulse_tail}Typical flux pulse distortion measurement for $\Phi_L$. The phase error $\delta \varphi$ versus the delay time $\tau_d$ after a large flux pulse $Z_0$ is shown. The orange line represents the fitting curve to perform correction. The blue dots show the phase error after the distortion correction. The insert figure shows the Ramsey-like pulse sequence for detecting the pulse distortion as a phase error.}
\end{figure}

Flux pulse distortion is a common concern for the control of frequency tunable superconducting qubits~\cite{rol2020time, sung2021realization}. For the $E_J$-tunable fluxonium considered here, there are two synchronously applied flux pulses with large amplitudes that need to be corrected. We use the qubit phase error as an $\insitu$ detector for the flux pulse distortion measurement and compensation, similar to that discussed in Refs.~\cite{barends2014superconducting, bao2022fluxonium}. The pulse sequence is shown in the insert of \autoref{fig:Zpulse_tail}. After the falling edge of a large and long flux pulse $Z_0$, a Ramsey like experiment is performed to trace the accumulated qubit phase under the following distorted flux pulse tail. During the qubit phase accumulation duration $\tau_p$, the qubit is offset to a flux sensitive point by a probing flux pulse $Z_p$. For the probing pulse, we use the combination of flux pulses in $\Phi_J$ and $\Phi_L$ with the amplitude $Z_{p,J}$ and $Z_{p,L}$ to shift qubit away from the sweet spot after the crosstalk correction. The qubit is always idled at the readout point in the absence of any flux pulse. Note that we measure and compensate the two fast flux pulses $Z_J$ and $Z_L$ independently, during which the initial flux pulse $Z_0$ is applied through $Z_J$ or $Z_L$ correspondingly. 

As shown by the line in \autoref{fig:Zpulse_tail}, the phase error data (for $Z_L$) can be well fitted with a three-component exponential function. The settling times and settling amplitudes for $Z_L$ are \{60.5, 420.8, 1533.7\}~ns and \{-2.87\%,  -1.52\%,  -1.06\%\}, respectively. For $Z_J$ the parameters are \{30.2, 113.2, 869.0\}~ns and \{-4.92\%,  -2.40\%,  -1.99\%\}. The flux pulses in $Z_L$/$Z_L$ are pre-distorted at the generator according to the above parameters using corresponding impulse response digital filters. In \autoref{fig:Zpulse_tail}, the data in blue dots shows the phase error after the pulse distortion correction for $Z_L$.

\begin{figure}
    \includegraphics[width=0.5\textwidth]{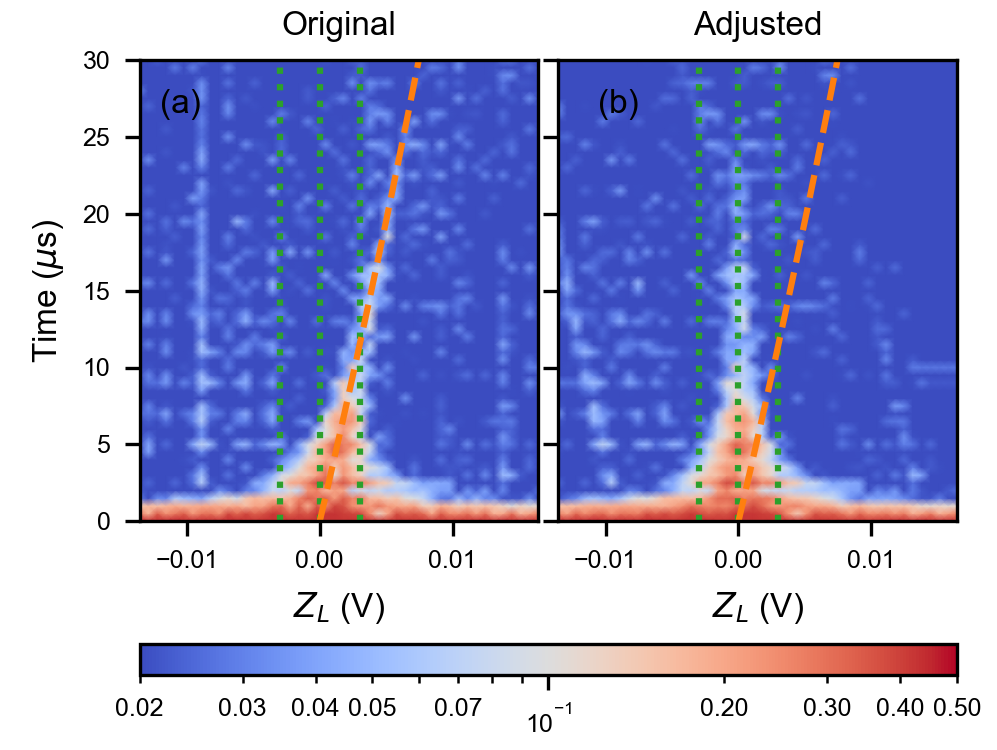}
    \caption{\label{fig:pulse_distortion}(a) Typical experimental data of the $T_2$ spin-echo measurements that shows $P_1$ envelop values decay profiles vs $Z_L$ at a specific $Z_J=-0.48$ V (corresponding to $\Phi_J/\Phi_0=0.239$). The $P_1$ decay profiles are tilted likely because of the residue flux distortion on top of the flux crosstalk. The dashed orange line is fit to the maximum $P_1$ of each horizontal trace to indicate the tilt. The dashed green vertical lines are guide for eyes. (b) The shifted $P_1$ along $Z_L$ axis for each time-length after fixing the tilt, making the orange line in (a) vertical. The dashed orange line and dashed green lines are the same as those in (a). }
\end{figure}

Even though we carefully correct the flux pulse distortion independently, in the decoherence measurements, we find another issue that is also induced by pulse distortion when the two flux pulses are combined applied to eliminate the crosstalk. Typical raw data of the $T_2$ spin-echo measurements along $\Phi_L$ is shown in \autoref{fig:pulse_distortion}(a). The $P_1$ decay profiles versus $Z_L$ at a specific $Z_J=-0.48$~V (corresponding to $\Phi_J/\Phi_0$=0.239) are shown. We find the $P_1(t)$ decay profiles are tilted at specific combinations of voltages $Z_L$ and $Z_J$ for two flux pulses. We know that at the sweet spot the $T_2$ should be the longest and the maximum $P_1$ point should be on a vertical line in the figure. But this line is obviously tilted, starting from the bottom center and ending at the top right, as indicated by the dashed orange line which fits the maximum $P_1$ at each delay. This implies that the combination of pulses $Z_L$ and $Z_J$ may have residue tails, especially when the amplitude of $Z_J$ becomes large ($|Z_J| \gtrsim 0.4$~V). To correct the pulse distortion induced $P_1$ tilt, we apply a $Z_L$-shift at each time-delay to ensure the maxima of $P_1(t)$ are always at the sweet spot ($Z_L = 0$). The amplitude of the shift follows an exponential decay, which is a common shape of pulse tails defined as $Z_c=a(1 - \exp(b  t))$. We use linear interpolation to determine the corrected $P_1'(Z_L, t)=P_1(Z_L+Z_c,t)$. After applying this data correction, the adjusted $P_1$ decay profiles are shown in \autoref{fig:pulse_distortion}(b).

It is worth mentioning that this behavior only affects $P_1$ points with long delay time compared to $T_2$, but the extracted $T_2$ from fitting to a decay are mainly determined by the decay of a much shorter time scale. So even without the extra data correction for pulse distortion, the extracted $T_2$ and flux noise do not change much. For example, we compare the extracted flux noise $A_L$ and $A_J$ (discussed in the following section) with and without data correction and only find a $5\%$ difference. 

\section{Additional decoherence data}
\begin{figure}
    \includegraphics[width=1.0\textwidth]{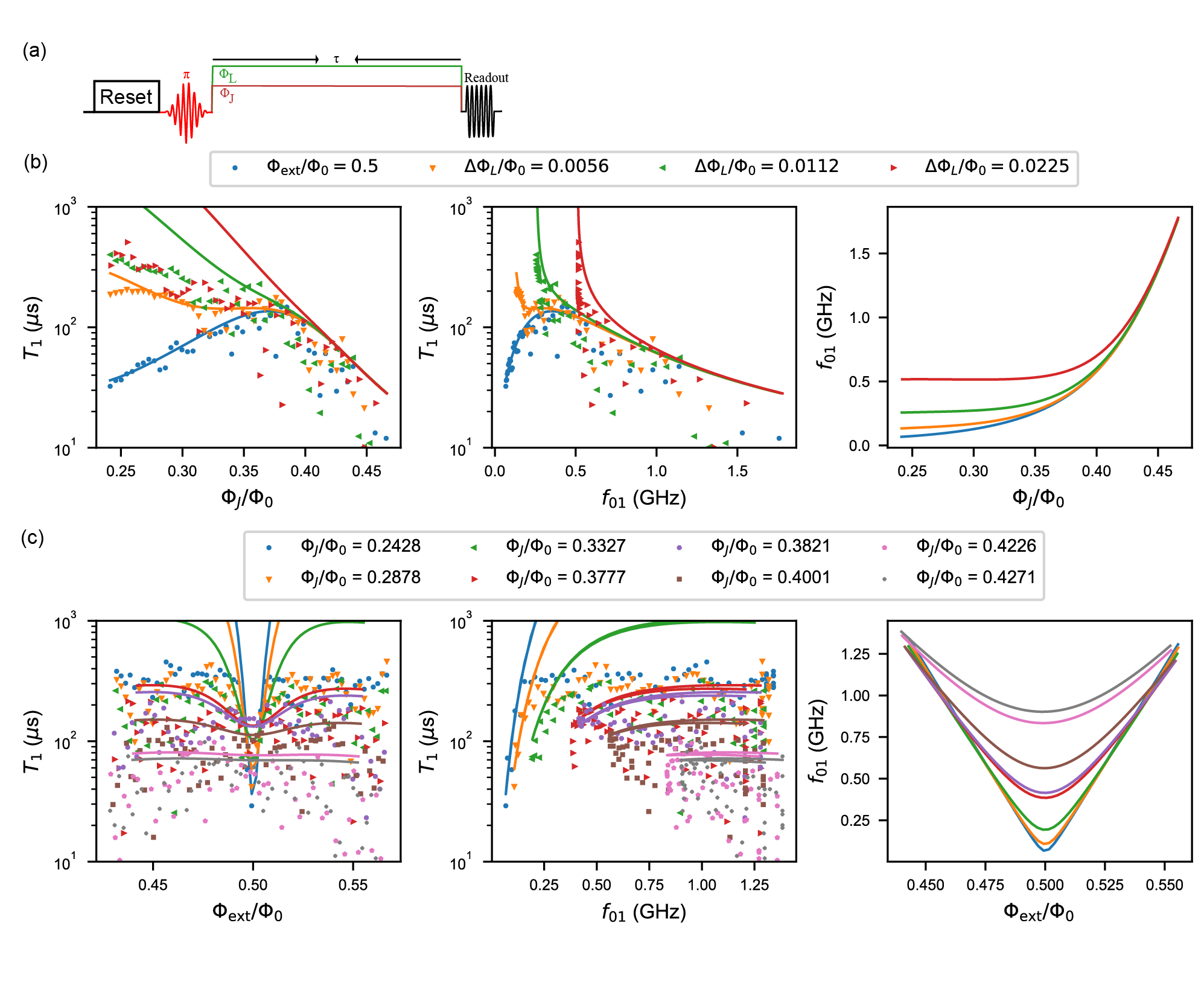}
    \caption{\label{fig:t1_vs_Phi_L}(a) Pulse sequence for measuring $T_1$ versus the two flux controls. (b) $T_1$ data along the $\Phi_J$ axis with the qubit at its $\Phiext/\Phi_0=0.5$ sweet spot as well as a few difference flux offset $\Delta\Phi_L$ away from the sweet spot. The left and middle panels are the same data set plotted versus $\Phi_J$ and qubit frequency $f_{01}$, respectively. The solid lines are calculated $T_1$ values using a model in combination of dielectric loss and loss due to $1/f$ flux noise. The right panel shows the dependence of qubit frequency on $\Phi_J$. (c) $T_1$ versus d $\Phiext$ around the $\Phiext/\Phi_0=0.5$ sweet spot at several $\Phi_J$ values together with the calculated values from the model are plotted versus $\Phiext$ (left panel) and qubit frequency $f_{01}$ (middle panel). The right panel shows the dependence of qubit frequency on $\Phiext$.}
\end{figure}

After calibrating the qubit parameters and their tuning range, we perform standard characterization of qubit decoherence in parameter regimes that are the most relevant for a fluxonium qubit. 
The pulse sequence for measuring the qubit energy relaxation using the combination of the two flux pulses is shown in \autoref{fig:t1_vs_Phi_L}(a). After reset, the qubit is prepared at its excited state with a $\pi$-pulse. With the external fluxes adiabatically actuated to different values with a varying duration $\tau_p$ for qubit relaxation but remained at the same levels for qubit reset, excitation, and readout, we are able to measure the qubit decoherence in a 2-dimensional parameter space spanning a large range of $\Phi_J$ and $\phiext$ with qubit operations calibrated only on a single point. 

For clarity, the $T_1$ data shown in Fig.~2(a) of the main text together with calculated values from the model are re-plotted in \autoref{fig:t1_vs_Phi_L}(b). This data set is measured along the $\Phi_J$ axis with $\Phiext/\Phi_0=0.5$ as well as a few difference flux offset $\Delta\Phi_L$ away from $\Phiext/\Phi_0=0.5$. We plot the $T_1$ data versus $\Phi_J$ (left panel) and qubit frequency $f_{01}$ (middle panel) and the corresponding the qubit frequency versus $\Phi_J$ curves are shown in the right panel. 

We also show $T_1$ data measured along the $\Phiext$ axis with fixed $\Phi_J$ in a $\Phi_0\pm 0.075\Phi_0$ range at several constant values of $\Phi_J$. As is shown in~\autoref{fig:t1_vs_Phi_L}(c), the data together with calculated values from the model are plotted versus $\Phiext$ (left panel) and qubit frequency $f_{01}$ (middle panel). The qubit frequency versus $\Phiext$ curves are shown in the right panel. For $\Phi_J < 0.38\Phi_0$ when the qubit frequency at the sweet spot is below 400~$\MHz$, we observe a sharp increase of $T_1$ when the qubit is moved away from its $\phiext = \Phi_0/2$ sweet spot, which is best described by flux noise. When $\Phi_J \gtrsim 0.38\Phi_0$, this $T_1$ increase is more gentle, which is consistent with dielectric loss. We note that unlike some demonstrations of heavy fluxonium where $T_1$ can reach a few milliseconds due to the small transition matrix element $\abs{\mel{0}{\hat{\varphi}}{1}}$ away from the sweet spot, we observe a limiting $T_1$ of around 400~$\mu$s. We hypothesize that this $T_1$ limit is imposed by the coupling to spurious modes in the environment.

\section{Dephasing in the $E_J$-tunable fluxonium}
From the definition of the reduced flux in the Hamiltonian in \autoref{eq:phij_to_phi12}, the noises of these two variables are correlated. The dephasing rate in the spin-echo measurements can be written as~\cite{gustavsson2011noise,nguyen2019high,ithier2005decoherence}
\begin{equation}
    \Gamma_{\phi}^2  = \ln 2\left(
    \left(\frac{\partial\omega_{01}}{\partial \Phi_L}\right)^2A_L^2+
    \left(\frac{\partial\omega_{01}}{\partial \Phi_J}\right)^2A_J^2\right. \left.+\frac{\partial\omega_{01}}{\partial \Phi_L}\frac{\partial\omega_{01}}{\partial \Phi_J}2c_{LJ}A_LA_J
    \right), \label{eq:tphi_flux_noise_two_loop}
\end{equation}
where $A_L$ and $A_J$ are the amplitude of the $1/f$ flux noise spectrum $S_{\Phi_\beta}(\omega)=2\pi A_\beta^2/\omega$ for $\Phi_L$ and $\Phi_J$ correspondingly, and $c_{LJ}$ is the correlation factor. The fit to the spin-echo measurements is shown in Fig.~2(b) of the main text, and the extracted parameters are $A_J=7.6~\mu\Phi_0$, $A_L=12.0~\mu\Phi_0$, and $c_{LJ}=0.51$. We can see the model and the data match quite well except that there could be a small offset in $\Phi_L$ between the model and data. This offset is estimated to be $\sim 0.0005\Phi_0$.

In our qubit design, the two loops share a very short section of a wire. So the correlation between the flux in two loops $\Phi_1$ and $\Phi_2$ can be assumed zero. Consequently, the correlation between $\Phi_L$ and $\Phi_J$ can be directly computed as~\cite{gustavsson2011noise,koch2007model}
\begin{align}
    c_{LJ} = \frac{1}{2}\frac{A_J}{A_L}. \label{eq:c_lj_zero_12_correlation}
\end{align}
The $c_{LJ}$ calculated from \autoref{eq:c_lj_zero_12_correlation} and $A_L$ and $A_J$ values is 0.32, smaller than the $c_{LJ} = 0.51$ extracted from fitting. This agreement is reasonable considering the simplicity of the model.

\section{Dissipation in the $E_J$-tunable fluxonium}
The dissipation due to the flux noise in the $E_J$-tunable fluxonium is similar to that in a traditional fluxonium~\cite{nguyen2019high} except that the single flux noise source should be replaced by two with a correlation factor. Following the same manner as deriving the dephasing model in the previous section, we have~\cite{gustavsson2011noise,nguyen2019high,ithier2005decoherence,schoelkopf2003qubits}
\begin{align}
    \Gamma_1  = & \frac{1}{\hbar^2}\left(
    V_L^2 S_L+
    V_J^2 S_J+V_LV_J2c_{LJ}\sqrt{S_L S_J}
    \right), \label{eq:t1_flux_noise_two_loop}
\end{align}
where $V_\beta=\left|\mel**{0}{\frac{\partial H}{\partial \Phi_\beta}}{1}\right|$ and $S_\beta = S_{\Phi_\beta}(\omega_{01})$ is the flux noise spectral density of $\Phi_\beta$ at the qubit frequency with $\beta = L \text{ or }J$. In the region of $\Phi_\beta$ where we perform measurements on our device, we verify that $V_J$ is always small compared with $V_L$. In this case, \autoref{eq:t1_flux_noise_two_loop} reduced to the relaxation due to a single flux noise source on $\Phi_L$, used in the main text.

\section{Noise spectral density data processing}
We extract the two-sided noise spectrum by a joint fit to the relaxation starting from the ground and excited state respectively. In this way, we can reliably estimate two key parameters, the relaxation time $T_1$ and stray population in equilibrium $p_\text{stray}$. This method is neither sensitive to readout errors nor the pulse distortions in long and large amplitude flux pulses. The symmetric and anti-symmetric noise spectra $S^+(\omega)$ and $S^-(\omega)$ in a form of flux noise are calculated using~\cite{quintana2017observation}
\begin{align}
    S^+(\omega) & =\frac{(2e)^2L^2}{T_1\left|\mel{0}{\hat{\varphi}}{1}\right|^2}, \label{eq:def_s+} \\
    S^-(\omega) & =(1-2\pstray)S^+(\omega). \label{eq:def_s-}
\end{align}
We note that at large flux pulse amplitudes (($|Z_J| \gtrsim 0.4$~V)), we observe that the relaxation deviates from an exponential decay at long decay time ($\tau \gtrsim 300~\mu$s). Such behavior is similar to what we observe in the $T_2$ data (see \autoref{fig:pulse_distortion}) but the tilted signal is the TLS traces. Therefore, this problem is probably due to uncompensated pulse distortions thus can be corrected by applying a shift on $Z_L$ at different time decay values. Here, we use data with $\tau \leq 300~\mu$s to process the noise spectra, where no pulse distortion problem is observed. We additionally note that even if we include all the delay time data with $\tau$ up to 1.5~ms, the obtained $T_1$ and $\pstray$ from the fit remained almost unchanged because the delay times are chosen evenly on a log scale where the majority of the data points are for catching the initial part of the relaxation. 

The effective temperature is estimated from $\pstray$ using $\exp(-\hbar \omega_{01}/k_B\Teff)=\pstray/(1-\pstray)$.

\section{Dielectric loss due to charge TLS}
Dielectric loss due to electric-dipole interaction with TLS was studied quite extensively in phase qubits~\cite{martinis2005decoherence} and transmons~\cite{barends2013coherent}. However, those previous works only consider the positive side of the noise spectrum for relaxation as the qubit frequency satisfies $\hbar\omega_{01}\gg k_B \Teff$.

Here, we extend those studies to low frequency qubits where the TLS is assumed to be in thermal equilibrium. The Fermi's golden rule becomes:
\begin{equation}
    \Gamma_{ij} = \frac{2\pi}{\hbar}\int_{\psi_{\text{TLS}}}\int D(p)\abs{\mel{j\psi_{\text{TLS}}}{H_{\text{int}}}{i\psi_{\text{eq}}}}^2 dp d\psi_{\text{TLS}}, \label{eq:Fermi}
\end{equation}
where $\ket{\psi_{\text{TLS}}} = \cos\frac{\theta'}{2}\ket{g}+e^{i\phi}\sin\frac{\theta'}{2}\ket{e}$ is an arbitrary TLS state, $\ket{\psi_{\text{eq}}} = \frac{1}{\sqrt{Z}}\ket{g}+\frac{\exp(-E/2k_B\Teff)}{\sqrt{Z}}\ket{e}$ is a pure state for representing the TLS in thermal equilibrium with $Z = 1+ e^{-E/k_B\Teff}$, and $D(p)$ is the TLS density of states with respect to the TLS dipole moment $p$. The qubit-TLS interaction Hamiltonian can be written as
\begin{align}
    H_\text{int} &= \frac{\hat V \hat p}{x}\cos \eta \nonumber \\
    &=\frac{i\hbar p_\text{max}}{2\varphi_0 x C}\left\langle {0}\left|{\frac{\partial}{\partial \varphi}}\right|{1}\right\rangle \cos\eta \sin\theta (\ket{0e}\bra{1g}-\ket{1g}\bra{0e}) \nonumber \\
    &= \frac{i \hbar p}{2\varphi_0 x C}\left\langle {0}\left|{\frac{\partial}{\partial \varphi}}\right|{1}\right\rangle(\ket{0e}\bra{1g}-\ket{1g}\bra{0e})
\end{align}
where we have used $\hat V = \frac{i\hbar\partial}{\varphi_0 C \partial \varphi}$ and $\hat p = p_\text{max} \frac{\ket{L}\bra{L}-|R\rangle\langle R|}{2}$; $\eta$ is the angle between the TLS dipole moment and the qubit electric field, $\theta = \Delta_0/\Delta$ is the transformation angle between the location states $\ket{L, R}$ and energy eigenstates $\ket{g,e}$ of the TLS, $x$ is a normalized distance between the qubit electrodes, and $C$ is the qubit capacitance. We note that $p = p_\text{max}\cos \eta \sin \theta$ is defined as the effective dipole moment (scalar) of the TLS measured by the transverse interaction with the qubit.

Following Ref.~\cite{martinis2005decoherence, muller2019towards}, the TLS density of states has an approximately uniform distribution over frequency and a distribution over the effective dipole moment given by $D(p) = \frac{d^2 N}{d E d p} = \rho_0 \sqrt{1-p^2/p_\text{max}^2}/p$, assuming the TLS asymmetry energy $\Delta$ has a uniform distribution and the tunneling energy $\Delta_0$ has a distribution proportional to $1/\Delta_0$. The integral in \autoref{eq:Fermi} can be evaluated using $\int_{\psi_{\text{TLS}}}d\psi_{\text{TLS}} = \frac{1}{4\pi} \int\sin\theta' d\theta' d\phi$. We obtain
\begin{align}
    \Gamma_{\downarrow} &= \frac{(2e)^2\tan\delta_C}{\hbar C} \left|\left\langle {0}\left|{\frac{\partial}{\partial \varphi}}\right|{1}\right\rangle \right|^2 \frac{2}{1+\exp(-\frac{\hbar\omega_{01}}{k_B \Teff})}, \nonumber \\
    \Gamma_{\uparrow} &= \frac{(2e)^2\tan\delta_C}{\hbar C} \left|\left\langle {0}\left|{\frac{\partial}{\partial \varphi}}\right|{1}\right\rangle \right|^2 \frac{2\exp(-\frac{\hbar\omega_{01}}{k_B \Teff})}{1+\exp(-\frac{\hbar\omega_{01}}{k_B \Teff})},
\end{align}
where the dielectric loss tangent is defined as
\begin{equation}
    \tan\delta_C = \frac{\pi\rho_0}{24(2e)^2C}\left(\frac{\hbar p_\text{max}}{\varphi_0 x}\right)^2.
\end{equation}
The relaxation rates obey detailed balance $\Gamma_\downarrow/\Gamma_\uparrow = \exp(\frac{\hbar \omega_{01}}{k_B \Teff})$, and we find that the total relaxation rate $\Gamma_1 = 1/T_1 = \Gamma_\downarrow + \Gamma_\uparrow$ is a temperature independent quantity. We have
\begin{align}
    \Gamma_1 &= \frac{2(2e)^2}{\hbar C}\tan\delta_C \left|\left\langle {0}\left|{\frac{\partial}{\partial \varphi}}\right|{1}\right\rangle \right|^2, \nonumber \\
    &= \frac{\hbar \omega_{01}^{2}}{4 E_C} \tan\delta_C \abs{\mel{0}{\hat{\varphi}}{1}}^2,
\end{align}
by rewriting the formula using the transition matrix element of the phase operator. Our result is consistent with that derived using Bloch-Redfield theory~\cite{shnirman2005low}.

\section{TLS spectroscopy}
\begin{figure}
    \includegraphics[width=0.7\textwidth]{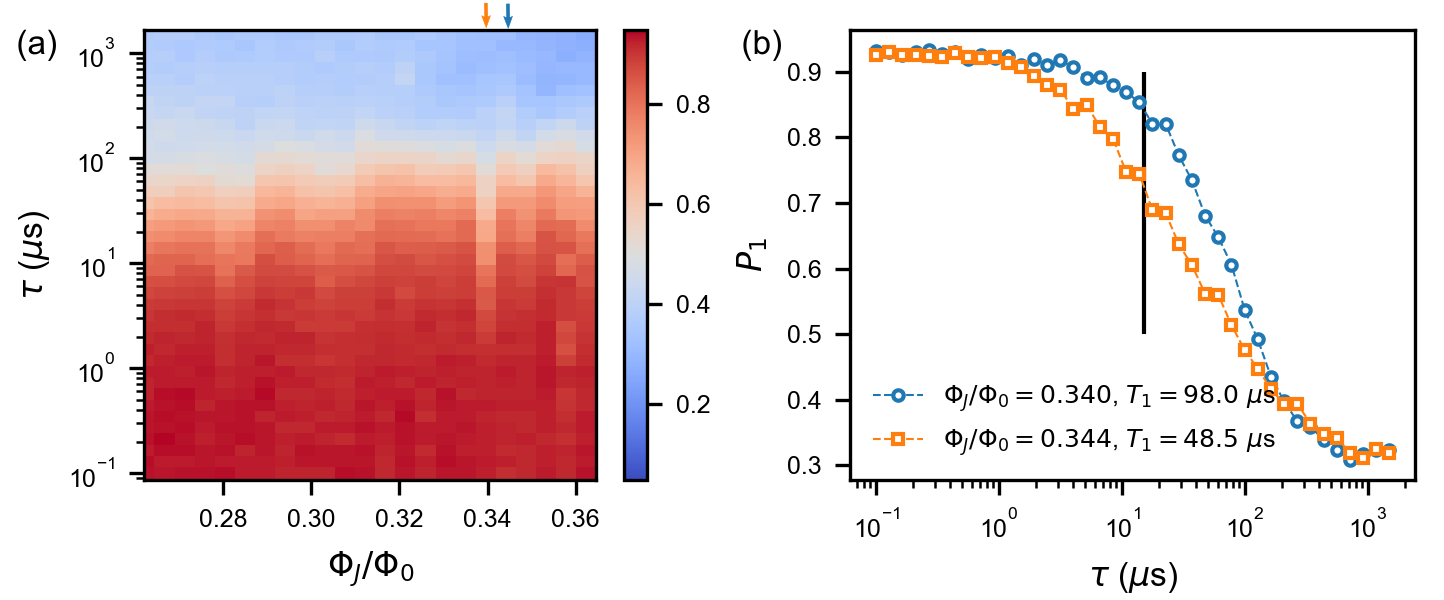}
    \caption{\label{fig:tls_data_process}(a) The experimental data of $T_1$ decay profiles measured along the $\Phiext/\Phi_0=0.5$ points with $\Phi_J$ varying from $0.26\Phi_0$ to $0.36\Phi_0$. The orange arrow marks a fast decay profile induced by a TLS. (b) $T_1$ decay profiles plotted together for comparison corresponding to the two arrow markers in (a). The black vertical cut indicates $\tau$ = 15$~\mu$s. The $P_1$ values at the cut can be a representative of the $T_1$ value.}    
\end{figure}

When the qubit is on resonance with a TLS, it undergoes a faster energy decay. Qubit-TLS coupling information can be extracted by mapping the qubit $T_1$ versus a qubit-frequency tuning range~\cite{grabovskij2012strain, klimov2018fluctuations, burnett2019decoherence, bejanin2021interacting, carroll2021dynamics}. To simplify and speed up the measurement, we pick a single $P_1$ at an specific delay in the qubit relaxation process to represent $T_1$. In \autoref{fig:tls_data_process}(a), we show the raw data of $T_1$ decay profile versus $\Phi_J$ with the qubit at the $\Phiext=\Phi_0/2$ sweet spot. When the qubit is on resonance with a TLS, as marked by the orange arrow, the qubit relaxes faster with a short $T_1$. For comparison, we plot the TLS induced fast decay profile (marked by the orange arrow) and a decay profile one $\Phi_J$ step away (marked by the blue arrow) in \autoref{fig:tls_data_process}(b). Fitting to a single exponential decay function, we have $T_1$ = 48~$\mu$s and $T_1$ = 98~$\mu$s respectively. Comparing with the $P_1$ values at $\tau= 15~\mu$s, as labeled by the vertical black line in \autoref{fig:tls_data_process}(b), the relative value of $P_1$ is a good representative of the $T_1$ value. With this simplified measurement, we map $P_1$ at delay $\tau= 15~\mu$s versus a large range of two flux controls $\Phi_J$ and $\Phi_L$. The result is shown in Fig.~4 of the main text. TLS-induced $P_1$-dips are in very good agreement with contours at constant qubit frequencies. 

\section{Simulation of qubit $T_1$ in a TLS defect bath}

We perform Monte Carlo simulations of the qubit in a bath of TLS defects. The contribution of each TLS defect to the qubit relaxation rate $\Gamma_1$ is given by  $\Gamma_1=\frac{2g^2\Gamma}{\Gamma^2+\Delta^2}$, a single Lorentzian peak~\cite{barends2013coherent} where $\Gamma=\Gamma_{2,Q}+\Gamma_{2,D}$. The TLS defects follow the standard tunneling model, where the asymmetry energy $\Delta$ is assumed to have a uniform distribution and the tunneling matrix element $\Delta_0$ has a distribution proportional to $1/\Delta_0$. Considering the relevant range of qubit frequency, the maximum values of $\Delta$ and $\Delta_0$ are both set to $3~\mathrm{GHz}\times h$ and the minimum values are $10^{-4}$ times the maxima. The qubit-TLS coupling $g$ depends on the dipole moment and the direction of the electric field of the qubit at the specific TLS location. For simplicity, we assume the charge matrix element $\mel{0}{\hat{n}}{1}$ of the qubit is proportional to the electric field. Following the study in the phase qubit~\cite{martinis2005decoherence}, the coupling can be written as  $g=|\mel{0}{\hat{n}}{1}|S_{\mathrm{max}}\cos\eta\sin\theta$ where $\eta$ is the angle between the TLS dipole moment and the electric field of the qubit and $\tan\theta=\Delta_0/\Delta$. We set $S_{\mathrm{max}}$ to $0.5~\mathrm{MHz}\times h$ to match our data. The $T_1$ and $T_2$ of the qubit without TLS defects are set to 1~ms and 50~$\mu$s. The $T_2$ of the TLS defects is assumed to have a uniform distribution between $5$-$100$~ns. We generate $10^4$ random TLS defects according to the above conditions to simulate the relaxation of the qubit at different frequencies.

\bibliography{ref}